\documentclass[journal=apchd5,manuscript=article]{achemso}
\usepackage{amssymb} 
\usepackage{color}
\usepackage[version=3]{mhchem} 
\usepackage{epstopdf} 

\usepackage{tabularx}
\usepackage{booktabs}
\newcommand{\tableheadline}[1]{\multicolumn{1}{c}{\small}}
\usepackage{caption}
\usepackage{subfig}  

\DeclareGraphicsExtensions{.pdf,.eps,.eps,.jpg,.mps}

	\title{Thermometric Calibration of the Ultrafast Relaxation Dynamics in Plasmonic Au Nanoparticles}
	\date{\today}
	
	\author{Marzia Ferrera}
	\affiliation{OptMatLab, Dipartimento di Fisica, Universit\`{a} di Genova, via Dodecaneso 33, I-16146 Genova, Italy} 
	\author{Giuseppe Della Valle}
	\affiliation{Dipartimento di Fisica, IFN-CNR, Politecnico di Milano, Piazza Leonardo da Vinci 32, I-20133 Milano, Italy}
	\author{Maria Sygletou}
	\affiliation{OptMatLab, Dipartimento di Fisica, Universit\`{a} di Genova, via Dodecaneso 33, I-16146 Genova, Italy} 
	\author{Michele Magnozzi}
	\affiliation{INFN, Sezione di Genova, via Dodecaneso 33, I-16146 Genova, Italy} 
	\alsoaffiliation{OptMatLab, Dipartimento di Fisica, Universit\`{a} di Genova, via Dodecaneso 33, I-16146 Genova, Italy} 
	\author{Daniele Catone}
	\affiliation{CNR-ISM, Division of Ultrafast Processes in Materials (FLASHit), Area della Ricerca di Roma Tor Vergata, Via del Fosso del Cavaliere, 100, I-00133 Rome, Italy} 
	\author{Patrick O'Keeffe}
	\affiliation{CNR-ISM, Division of Ultrafast Processes in Materials (FLASHit), Area della Ricerca di Roma 1, I-00015 Monterotondo Scalo, Italy} 
	\author{Alessandra Paladini}
	\affiliation{CNR-ISM, Division of Ultrafast Processes in Materials (FLASHit), Area della Ricerca di Roma 1, I-00015  Monterotondo Scalo, Italy} 
	\author{Francesco Toschi}
	\affiliation{CNR-ISM, Division of Ultrafast Processes in Materials (FLASHit), Area della Ricerca di Roma 1, I-00015  Monterotondo Scalo, Italy} 
		\author{Lorenzo Mattera}
	\affiliation{OptMatLab, Dipartimento di Fisica, Universit\`{a} di Genova, via Dodecaneso 33, I-16146 Genova, Italy} 
\author{Maurizio Canepa}
	\affiliation{OptMatLab, Dipartimento di Fisica, Universit\`{a} di Genova, via Dodecaneso 33, I-16146 Genova, Italy} 
	\author{Francesco Bisio}
	\affiliation{CNR-SPIN, C.so Perrone 24, I-16152 Genova, Italy}	
\email{francesco.bisio@spin.cnr.it}
	
\begin{document}

	\textcolor{red}{This article may be downloaded for personal use only. Any other use requires prior permission of the author and ACS Publishing.  This article appeared in (M. Ferrera, ACS Photonics 2020, 7, 4, 959–966) and may be found at (https://10.1021/acsphotonics.9b01605).}

\newpage

\begin{abstract}
The excitation of plasmonic nanoparticles by ultrashort laser pulses
sets in motion a complex ultrafast relaxation process involving the gradual re-equilibration of the system's electron gas, lattice and environment.
One of the major hurdles in studying these processes is the lack of {\em direct} measurements
of the dynamic temperature evolution of the system subcomponents.\\
We measured the dynamic optical response of ensembles of plasmonic Au nanoparticles following ultrashort-pulse
excitation and we compared it with the corresponding static optical response as a function of the increasing temperature
of the thermodynamic bath.
Evaluating the two sets of data, the optical fingerprints of equilibrium or off-equilibrium responses  could be clearly identified, allowing us to extract a dynamic thermometric calibration scale
of the relaxation process, yielding the experimental ultrafast temperature evolution
of the plasmonic particles as a function of time.
\end{abstract}

{\em keywords: Plasmonics, Gold Nanoparticles, Ultrafast Dynamics, Thermoplasmonics, Optical Spectroscopy}

\newpage

The irradiation of nanometric metallic particles (NPs) with ultrafast laser pulses 
sets in motion a time-dependent dynamics that affects the electron gas, the lattice and the local environment of the NPs
\cite{Voisin01,Link03,Hartland11,Hartland04,Jain06,Feldstein97,Baffou11,Baffou13}.
Depending on the characteristics of the exciting radiation (fluence, wavelength, pulse duration {\em etc.}), the NPs may exhibit 
radically-different time-dependent evolutions,  along which intriguing, off-equilibrium states of matter may occur
\cite{Taylor14,Herrmann14,Gonzalez16,Catone18}.
The typical relaxation pathway of  NPs lying 
within a thermodynamic bath at temperature $T_{bath}$ 
and impulsively excited by a radiation pulse at $\tau=0$ can be schematically divided in three distinct steps
\cite{Pustovalov16,Brongersma15}.
First, the radiation pulse induces an out-of-equilibrium electron population that
thermalizes {\em via} $e$-$e$ scattering at $T_e\gg T_{bath}$ within few 100-{\em fs} \cite{BisioPRB96};
then the energy exchange with the NP lattice {\em via} $e$-$ph$ coupling leads to the thermalization, on the few-$ps$ time scale, of the
NP lattice with the electron gas  at $T_e\!=\!T_l >T_{bath}$  and, finally, 
phonon-phonon interactions lead to the thermalization of the NPs with their environment ($T_e\!=\!T_l\!=\!T_{env}$)  over tens to hundreds of $ps$.
The energy dissipation to the thermodynamic bath determines the subsequent, slow asymptotic evolution 
of the overall temperature towards $T_{bath}.$

The process outlined above is however far from being straightforward.
The photon energy, the radiation fluence $F$, 
 the NP size \cite{Halte1999} 
and environment \cite{STOLL_jpcc_2015}, the occurrence of plasmonic resonances and hot-spots \cite{Manjavacas14,Harutyunyan15,Magnozzi_JPCC123_16943_2019} or the hot-electron injection from NPs to their environment \cite{Brongersma15,Zhang14,Saavedra_ACSPhoton2016} all play a role in influencing the actual relaxation pathway.
Additionally, the electron and lattice heat capacitances, the materials' dielectric functions \cite{Magnozzi2019,Ferrera_PRM2019}
and
the related-interface thermal resistances, exhibit an (often unexplored) temperature dependence. Last but not least, temperature-driven morphological effects can come into play \cite{Plech04} and the environment temperature $T_{env}$ may exhibit huge spatial gradients 
around the NPs on the $ps$ time scale. 
Given this, it is hardly surprising that understanding and modelling the ultrafast relaxation dynamics of impulsively-excited matter and of plasmonic NPs is a task 
that remains challenging to this day \cite{Juve_PRB2009,Strasser14,Brown16,Pustovalov14,Stoll2014,%
Labouret16,Saavedra_ACSPhoton2016,Brown_PhysRevLett2017,Zavelani_ACSPhoton,jermyn_transport_2019,Block_SciAdv2019}.\\
Experimentally,
the progresses in ultrafast optical and electronic spectroscopies have allowed, over the years, deeper and deeper insights in these complicated processes \cite{Guillet09,Wang15,Hobbs2018,lietard_electron_2018,Ortolani_PhysRevB99_2019,Block_SciAdv2019}, ultimately
promoting the emergence of the above-described general picture.
One crucial point is however still represented by the fact that, while $T_e$, $T_l$ and $T_{env}$ are basic ingredients of  theoretical models, their evaluation from experimental data is typically always {\em indirect} \cite{VandeBroek11,Plech_Nanoscale2017}.

In this work we report a simple experimental approach that may significantly
improve the quantitative understanding of ultrafast dynamics
in solid-state systems.
We investigated the optical response of ordered arrays of plasmonic Au NPs under two different experimental conditions.
First, we recorded their ultrafast optical response following pulsed-laser irradiation, by means of
pump-probe transient absorbance spectroscopy (TAS).
Then, we measured their {\em static} optical response as a function of variable thermodynamic-bath temperature, $T_{bath}$,
in a separate experiment.\\
In the former case, $T_e$, $T_l$ and $T_{env}$ are complicated functions of the delay time 
$\tau$ elapsed since the exciting pulse. 
In the latter case, by definition $T_e\!=\!T_l\!=\!T_{env}\!=\!T_{bath}$, where $T_{bath}$ is known.
Comparing the ultrafast response with the static
data, we were therefore able to ascertain the optical fingerprints of equilibrium or off-equilibrium behaviour and
experimentally estimate 
the delay time for for which the intraparticle equilibrium condition ($T_e = T_l$) is achieved.
From then onwards, we could follow the temperature-relaxation dynamics,
extracting an effective thermometric calibration scale that allowed us to assess
the ultrafast temperature evolution of the NP ensembles.\\

\section*{Experimental}

\begin{figure}[!htbp]
\includegraphics[width=8.0cm]{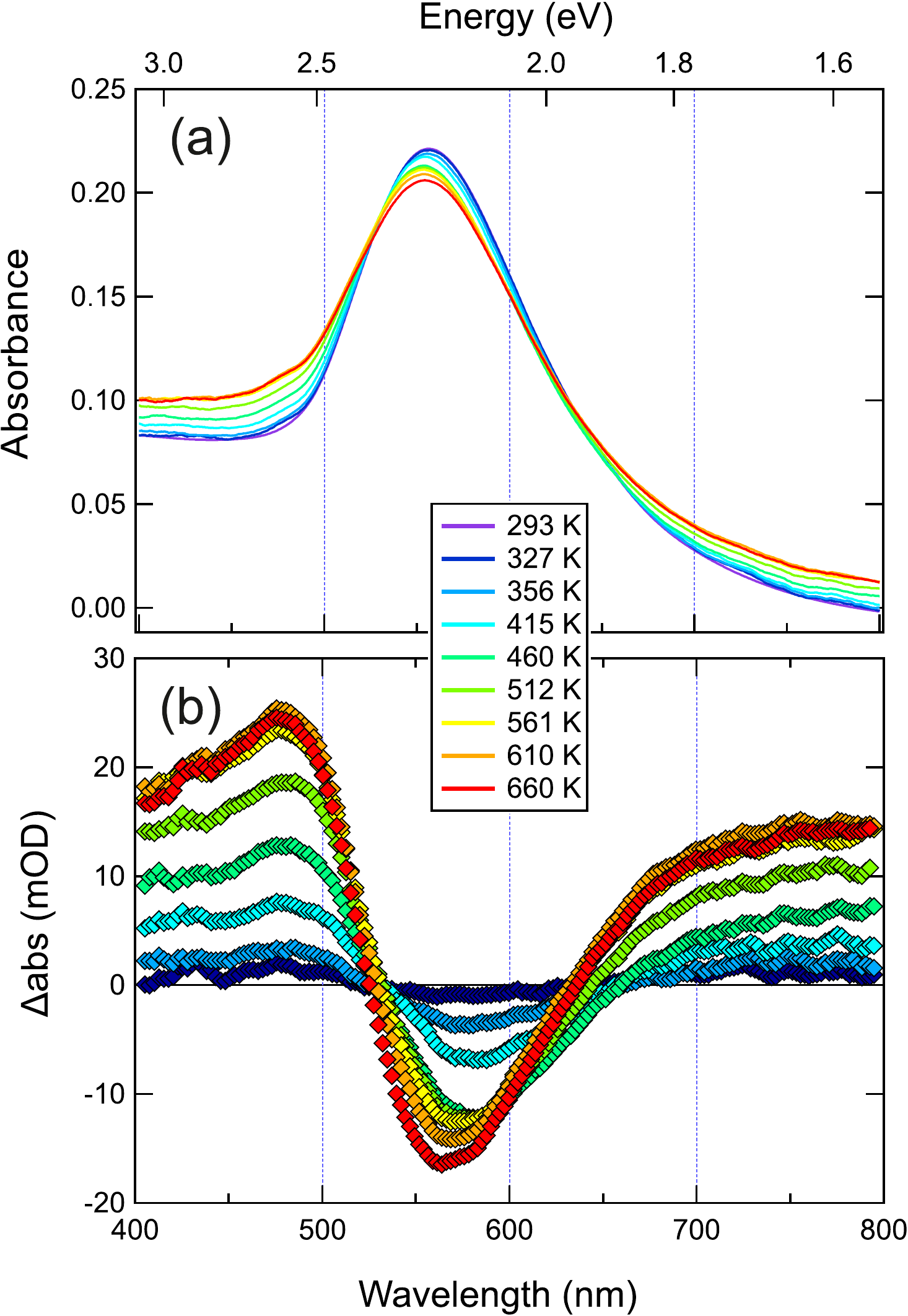}
\caption{Panel (a): experimental static absorbance spectra of Au NPs as a function of $T_{bath}$.
Panel (b): differential-absorbance spectra relative to room-temperature absorbance, calculated
from the data of panel (a).}
 \label{thermo}
\end{figure}

The samples consisted of 2D arrays of Au NPs fabricated by solid-state dewetting of 
ultrathin Au films onto a nanopatterned LiF(110) single crystal.
The sample fabrication procedures are described in detail in previous works
\cite{Anghinolfi_JPCC_2011,Zaccaria,Magnozzi_JPCC123_16943_2019,Sugawara_JVSTB23_443_2005}.
The system is composed of densely-packed Au NPs (areal density $10^3\pm 100 \:\mathrm{NP}/\mu\mathrm{m}^2$), with mean size around 20-25 nm, circular in-plane
cross section and short-range order, laid on a transparent LiF(110) substrate (Crystec Gmbh).
An atomic-force microscopy image of the system is reported in the Supporting Information (Figure S1).
The system exhibits a room-temperature (RT) localized surface plasmon resonance (LSPR) at $\lambda_{LSPR}\approx 560$ nm, as reported in Fig. \ref{thermo}(a) for 293 K.

The static optical response of the NPs as a function of $T_{bath}$ was assessed by means of 
temperature-dependent optical trasmittance spectroscopy
(400-800 nm range), performed in a home-designed high-vacuum (HV) vessel ($p<10^{-7}$ mbar).
White light from an incandescence W lamp was collimated, linearly polarized, and directed onto the sample
through a transparent stress-free optical viewport (see the Supporting Information (Figure S2)).
The transmitted light exited the HV chamber through another viewport and was collected by a spectrometer (Ocean Optics USB2000+).
The transmittance spectra were normalized to the bare-substrate response.
The lamp intensity decreases at both ends of the spectrum, causing a larger noise level at wavelengths below 450 nm and above 700 nm.
Temperature dependent spectra were collected during a decreasing-$T_{bath}$ ramp, starting from the maximum temperature $T_{bath}=660$ K
until RT, in order to limit any spurious influence of weakly-bound contaminants.
Upon approaching RT, liquid nitrogen cooling was applied to speed the temperature ramp and limit any unwanted effect
due to a prolonged permanence in HV.\\
The ultrafast, time-resolved measurements were performed by means of TAS.
The laser pulses used for exciting/heating and probing of the 2D arrays were produced by a single femtosecond laser system. This system consists of a Ti:Sapphire oscillator delivering 20 fs pulses with a frequency of 80 MHz, a part of whose output is used to seed a chirped pulse amplifier which in turn generates 4 mJ, 35 fs pulses centered at 800 nm with a repetition rate of 1 kHz.  An optical parametric amplifier (OPA) is then used to convert part of the amplifier output into tunable radiation. 
The output of the OPA at a wavelength of 410 nm is used as the pump pulse.
The experimental fluence on the sample was $F=(4\pm2)\;\mathrm{J}/\mathrm{m}^2$. 
The error stems from the uncertainty on the spot diameter dimension $(400 \pm 100)\; \mu\mathrm{m}$.
For the probe, white light supercontinuum (SC) pulses (350-800 nm) are generated by focusing 3 $\mu$J of the amplifier radiation into a rotating CaF$_2$ crystal.
The latter is used in a split-beam configuration in which 50\% of the white light passes through the sample while the remainder is used as a reference to account for pulse-to-pulse fluctuations in the white-light generation.
The delay between the two pulses is scanned by varying the optical path of the probe light \cite{Fratoddi17}.
The absence of damage following the laser irradiation was assessed by checking the stability of
the sample morphology and optical response. Variations of sample morphology were observed for laser fluences of $13\; \mathrm{J}/\mathrm{m}^2$ and above \cite{Magnozzi_JPCC123_16943_2019}.

In Fig. \ref{thermo}(a) we report a set of static absorbance spectra of the Au NPs as a function of $T_{bath}$;
the peak at $\lambda\approx 560$ nm is the fingerprint of the LSPR of the NP array, 
red shifted and broadened with respect to the LSPR of the individual NPs because of coupling effects and inhomogeneities in the array \cite{Zaccaria}.
As a function of increasing $T_{bath}$ we can observe that the LSPR absorbance peak broadens and weakens, 
due to the larger optical losses induced by the increased phonon population (plasmon bleaching, PB).
The full-width at half maximum of the LSPR peak evolves from 0.46 eV to 0.49 eV from RT to 660 K,
similar to previous observations \cite{Doremus_JChemPhys1964,YESHCHENKO2013275}.
To better highlight the temperature effects, in Fig.\ref{thermo}(b) we report the 
corresponding differential-absorbance spectra $\Delta\mathrm{Abs}$,
calculated as
$\Delta\mathrm{Abs}(T_{bath})=\mathrm{Abs}(T_{bath})-\mathrm{Abs}(293\; \mathrm{K})$.
The plasmon bleaching is apparent as the prominent negative feature in the center of the spectral range,
whereas positive wings (photo-induced absorption, PIA) arise from the LSPR broadening with increasing $T_{bath}$.
The PB and PIA decrease in magnitude with decreasing temperature, and
the PB peak position initially redshifts with increasing $T_{bath}$, then blueshifts from 415 K onwards.
The initial redshift of the PB peak followed by its blueshift are a consequence of the fact that the LSPR peak very slightly blueshifts for T up to 415 K, and then redshifts with increasing T.
We believe that the experimental trend of the PB peak reflects two conflicting behaviors, namely a redshift of the LSPR induced by the gradual broadening of the interband transitions of Au with increasing T, and a blueshift due to the temperature-induced shift of the plasma frequency to higher wavelengths \cite{Magnozzi2019}. More complex effects depending on the peculiar configuration of the sample, including thermal dilatation and subsequent change in the interparticle coupling, could also contribute to these subtle dispersive features of thermo-modulation spectra, whose precise explanation is beyond the scope of the present study.

\begin{figure}[!htbp]
\includegraphics[width=8.0cm]{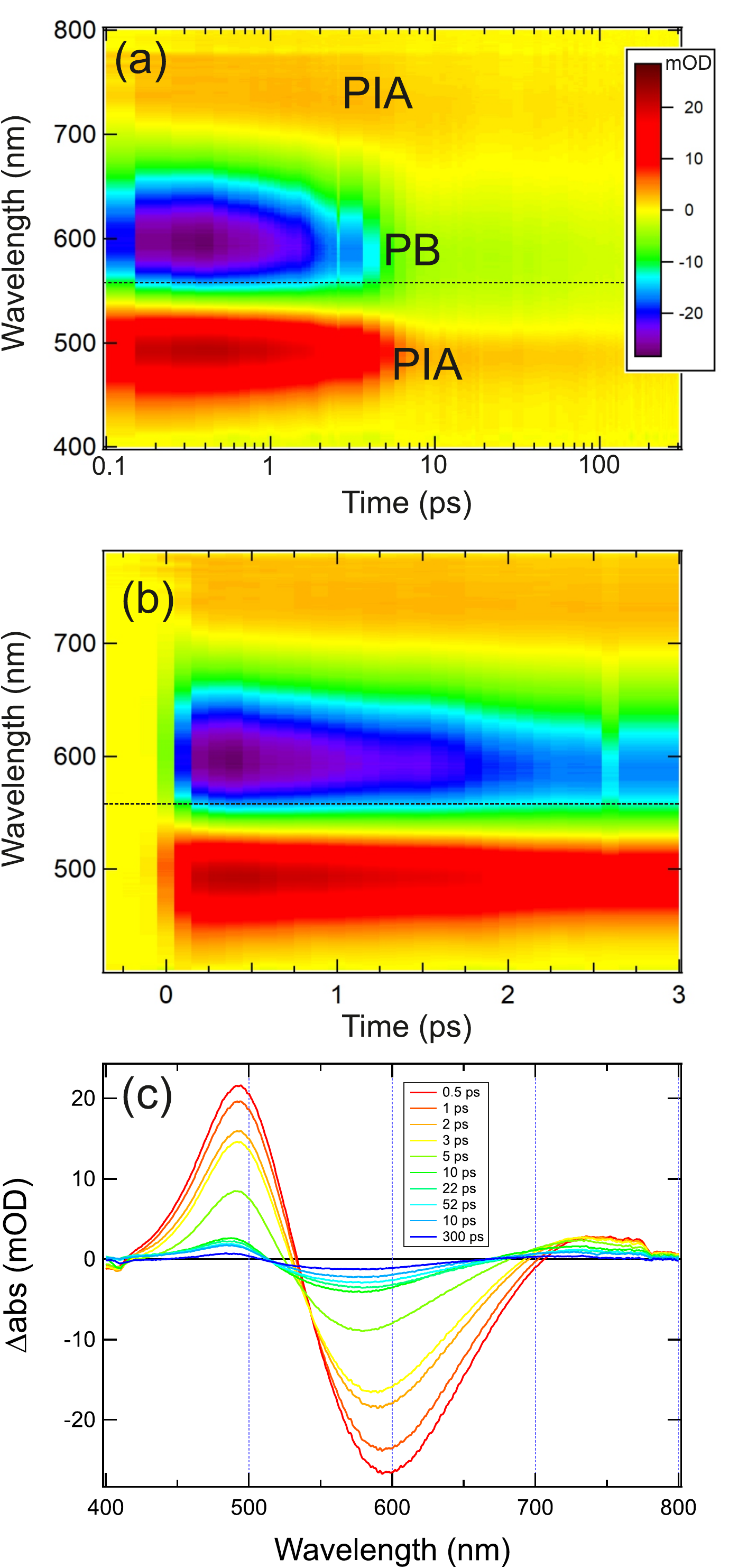}
\caption{Panels (a), (b): Femtosecond transient-absorbance map of Au NPs.
The vertical scale represents the probe wavelength, the horizontal scale the delay time $\tau$ elapsed since the
410-nm excitation pulse, on a logaritmic (a), and linear (b) scale, respectively.
Panel (c):  transient absorbance spectra as a function of time-delay $\tau$, extracted from the
map of panels (a), (b).}
  \label{TAS}
\end{figure}

In Fig. \ref{TAS}(a),(b) we report the results of TAS measurements performed on the same system.
On the horizontal scale is the time delay $\tau$, expressed in $ps$, and on the vertical scale, the probe wavelength.
The color scale represents the $\Delta\mathrm{Abs}$ of the system, expressed in mOD, as the difference between the absorbance of the excited and unperturbed system, respectively.
In Fig. \ref{TAS}(a) we report a broad time scan, up to $\tau\approx 300\; ps$, on a logaritmic time scale, whereas
in Fig. \ref{TAS}(b) we focus on the first few $ps$ of dynamics.
The horizontal dashed line represents the wavelength of the static LSPR maximum at RT.
The TAS data clearly show the PB feature at $\lambda\approx 595$ nm, accompanied
by the PIA on the wings of the LSPR.
After a sharp rise in the first few 100 {\em fs}, all the features fade in intensity with growing $\tau$,
in agreement with expectations \cite{Hartland04,Brown_PhysRevLett2017,Wang15} and the PB peak gradually blueshifts, inching closer to the static LSPR wavelength.\\
In Fig. \ref{TAS} (c), we report TA spectra ({\em i.e.} spectral cuts at selected $\tau$), ranging between 0.5 {\em ps} and 300 {\em ps}.
In those spectra, the PB is apparent as the prominent negative feature at $\lambda\approx 600$ nm, whereas
the PIA structures are found around $\lambda\approx 490$ nm (the most intense) and
at $\lambda>700$ nm.
Here, we observe that the PB peak monotonously blueshifts as a function of increasing delay.

\begin{figure}[!htbp]
\includegraphics[width=15.0cm]{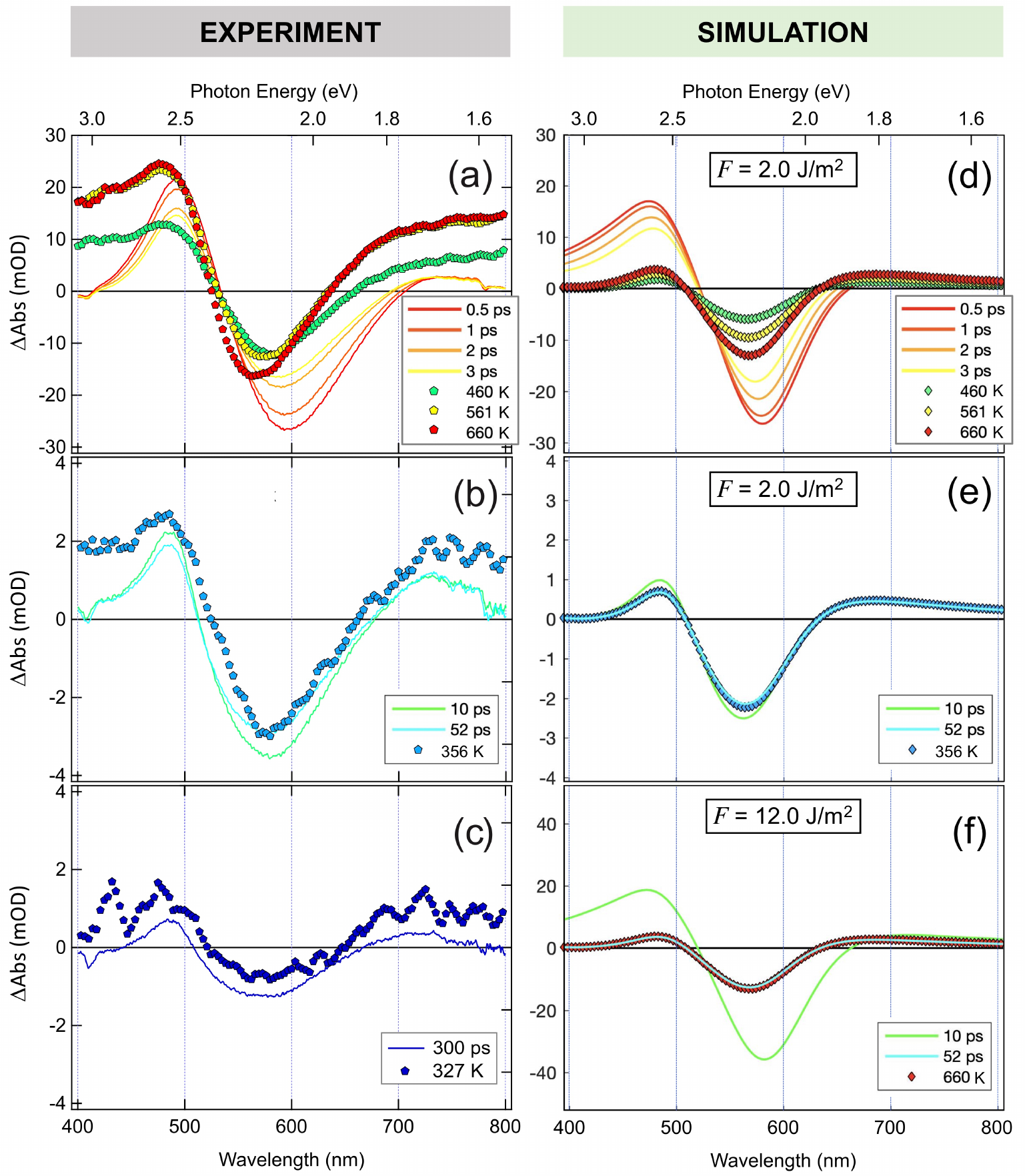}
\caption{Comparison between static $T_{bath}$-dependent $\Delta\mathrm{Abs}$ spectra (symbols) and delay-dependent dynamic TA spectra (lines).
In panels (a), (b) and (c) are reported data for gradually-decreasing $T_{bath}$ and gradually-increasing time delay $\tau$, respectively. Panels (d) and (e) show numerical simulations corresponding to the experiments detailed in panels (a) and (b), respectively. Panel (f) reports on the theoretical study case with a pump fluence close to the damage threshold.}
  \label{delays}
\end{figure}

\section*{Discussion}

Comparing the two sets of data of Fig. \ref{thermo} and \ref{TAS}, we can notice that the short-delay intensity of the PB and PIA features in TA 	spectra matches the static values of PB and PIA intensity retrieved for high-$T_{bath}$, and analogously so for
intermediate-delays/intermediate-$T_{bath}$ and long-delays/low-$T_{bath}$ data.
In Fig. \ref{delays} we therefore directly compare sets of $\Delta\mathrm{Abs}$ spectra recorded in the two different experiments.
In panel (a) we report short-delay TA spectra ($\tau=0.5$ {\em ps} to $\tau=3$ {\em ps}) 
along with high-$T_{bath}$  $\Delta\mathrm{Abs}$ spectra (460 K to 660 K), while
in panels (b), (c) we compare intermediate-delay TA spectra (22 {\em ps} and 52 {\em ps}) 
with static $\Delta\mathrm{Abs}(T_{bath}=356$ K),
and long-delay TA spectra ($\tau=300$ {\em ps}) with $\Delta\mathrm{Abs}(T_{bath}=327$ K), respectively.

The overall spectral shape of the $\Delta\mathrm{Abs}$ spectra is similar for all data, yet it is apparent
that none of the high-$T_{bath}$ {\em static} spectra can match the short-delay TAS data (Fig. \ref{delays}(a)).
In particular, large discrepancies in the spectral position of the PB peak and of the long-$\lambda$
zero crossing appear.
This clearly underscores the well-known fact that impulsive excitations at ultrashort delays do {\em not}
simply heat the NPs, but create out-of-equilibrium conditions whose optical response cannot be reproduced 
by any equilibrium counterpart \cite{Brown16}.
Increasing the delay and decreasing $T_{bath}$, the two sets of spectra gradually get more similar
(Fig. \ref{delays}(b), $\tau=22-52 \;ps$, $T_{bath}=356$ K), until
they almost overlap, within experimental uncertainty (Fig. \ref{delays}(c), $\tau=300 ps$, $T_{bath}=327$ K) .

\begin{figure}[!htbp]
\includegraphics[width=9.3cm]{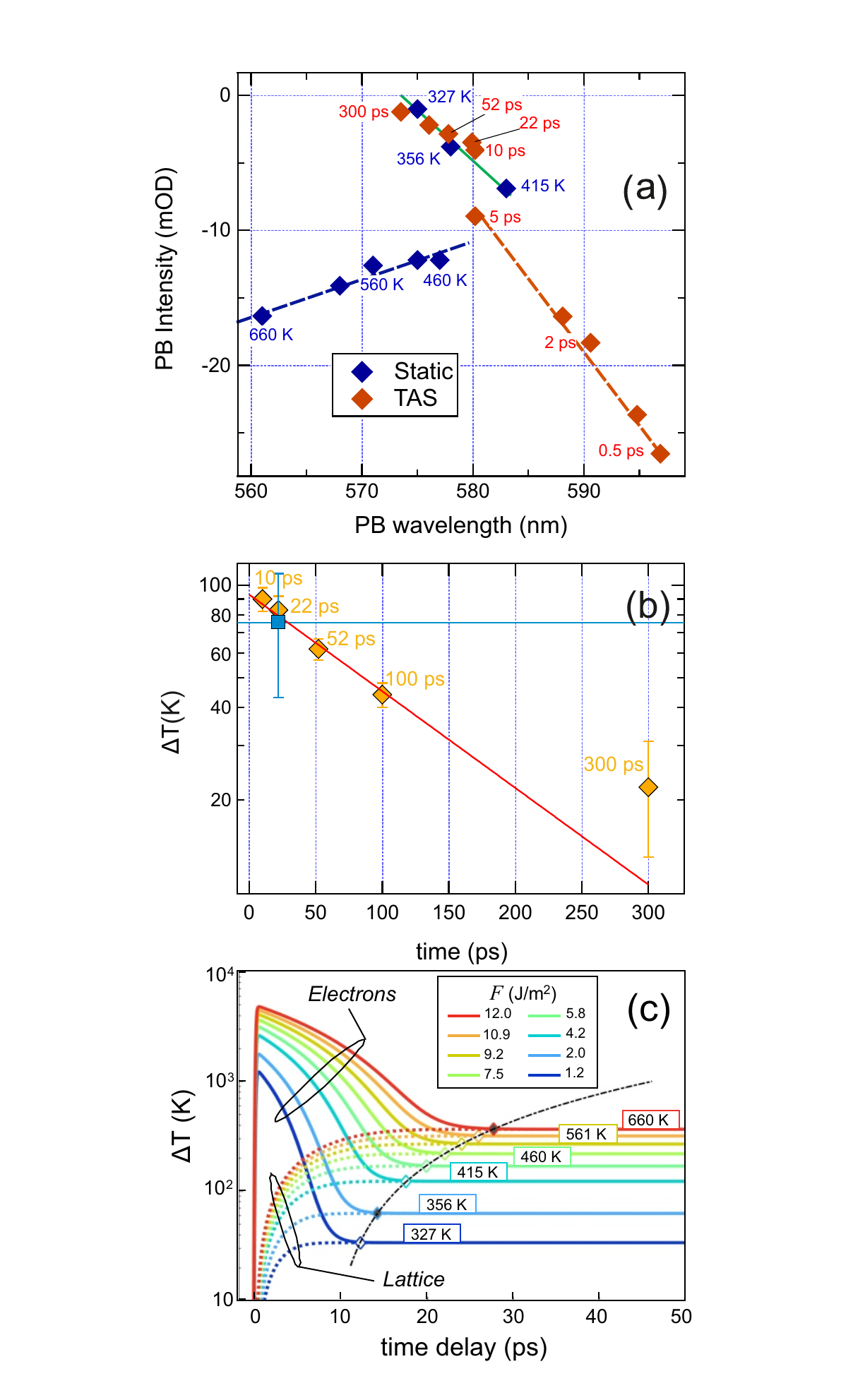}
\caption{Panel (a): PB-peak intensity {\em vs} PB-peak wavelength extracted from dynamics TA spectra as a function of
the delay time $\tau$ (red markers) or extracted from static $T_{bath}$-dependent data (blue markers).
The values of $\tau$ and $T_{bath}$ corresponding to each experimental point are reported in the graph.
The dashed lines are a guide to the eye. The continuous green line is an interpolation of the static data in the
(415-327) K $T_{bath}$ interval.
Panel (b): time dependence of the NP temperature extracted from the analysis of the data in panel (a) (orange symbols).
Estimated low-delay NP temperature according to Eq. \ref{eq:maxT} (blue symbol).
Exponential fit to the orange markers (red line). (c) Simulated electron (solid curves) and lattice (dashed curves) temperature dynamics  after excitation with fs-laser pulses of increasing pump fluence. Diamonds mark the time delays $\tau_{EQ}$ of electron-lattice equilibration, with the cases corresponding to the simulations of Fig.~\ref{delays}(d)-(f) highlighted by gray filling. Dash-dot black line is a fitting ($R^2 = 0.9999$) power law $\tau_{EQ}(\Delta T_{max}) = a\times (\Delta T_{max})^{2/3}+b$, with $a = 0.385$~{\em ps} K$^{-2/3}$, $b=8.193$~{\em ps.}} 
\label{summary}
\end{figure}

In order to quantify the gradual rapprochement of static and dynamic data, we report in Fig. \ref{summary}(a) the
evolution of the PB-peak intensity $vs$ its wavelength for different static $T_{bath}$ (blue symbols)
and delay time (red symbols) values (we do not report the experimental uncertainty for the sake of clarity).
The trajectories of PB-peak evolution (visually highlighted by the dashed lines in Fig. \ref{summary}(a)) 
lie very far apart from each other for high-$T_{bath}$/small-$\tau$ values, then
become closer for $\tau\rightarrow5$ $ps$ and $460\;\mathrm{K}>T_{bath}>415\;\mathrm{K}$, and
start overlapping  for $\tau\approx10$ $ps$ and
$T_{bath}\approx 356$ K.
From there onwards, they follow the same path (green line) until the
last recorded experimental points ($\tau=300$ $ps$, $T_{bath}=327$ K).

The increasing similarity between static and dynamic data for long-$\tau$/low-$T_{bath}$ is due to the achievement
of thermodynamic equilibrium within the impulsively-excited system.
Indeed, after few $ps$, $T_e=T_l$, and the
heat diffusion to the substrate becomes the only significant ongoing process 
\cite{Pustovalov14,DellaValle2012,Brongersma15}.
At this stage, since LiF exhibits very low variations in refractive index with temperature \cite{Li_JPVRD_1976}, the substrate heating is unlikely
to have a significant effect on the overall plasmonic response. 
Thus the dynamic response for $\tau \geq 10\; ps$ is expected to be
similar to the static response at aptly-chosen values of $T_{bath}>$ RT.
The small discontinuity in TAS data trend between the values at 5 {\em ps} and 10 {\em ps}
is ascribed to NP breathing modes \cite{Hodak_1998} (clearly absent in static data).

The good overlap between dynamic and static data suggests 
the possibility to exploit the static $T_{bath}$ data as an effective thermometric calibration for their dynamic counterpart.
In order to do this, one can perform a linear fit across the low-$T_{bath}$ static points (green line in Fig. \ref{summary}(a)) and
build an effective thermometric scale using the experimental static $T_{bath}$ values as reference.
Then, the experimental TA points can be projected onto the calibration line
and, based on the position of the projection point, the dynamic temperature can be estimated.
To do this, we chose to perform an orthogonal geometrical projection in the wavelength-intensity plane, 
meaning that the calibration temperature for a given TAS point corresponds to the value along the green line having the
shortest distance from the point itself.

The result of this analysis, {\em i.e.} the delay-dependent temperature difference $\Delta T(\tau)$, with respect to RT,
 is reported in Fig. \ref{summary}(b) (orange symbols).
Treating the temperature evolution along the calibration line ({\em i.e.} the green line in Fig. \ref{summary}(a)) as either a linear function, a polynomial function or an exponential curve,
respectively, yields
slightly different temperatures for the dynamic data, giving rise to the effective experimental uncertainty
reported in Fig. \ref{summary}(b).
The $\Delta T(\tau)$ decay can be well fitted by means
of an exponential function, as expected, yielding a time constant of $\alpha=(135\pm20)$ {\em ps} (red line).
The agreement is quite good, save for the last data point, where the experimental uncertainty
becomes larger due to the small magnitude of the experimental $\Delta$Abs spectra.

In order to independently test the validity of such a thermometric calibration, we can 
independently evaluate the maximum laser-induced temperature increase of the NPs, $\Delta T_{max}$
according to Ref. \citenum{Baffou11}.
Under the hypothesis that the electromagnetic energy absorbed by the NP from the light pulse is used to homogeneously heat the NPs,
that the dielectric function of Au is constant over the duration of the excitation pulse, and that no heat dissipation to the environment has occurred (an approximation valid within few tens of {\em ps} after the exciting pulse) $\Delta T_{max}$ can be expressed as: 

\begin{equation}
\label{eq:maxT}
\Delta T_{max}=\frac{\sigma_{abs}F}{V\rho_{Au}c_{Au}}
\end{equation}

where $V$ is the NP volume, $\rho_{Au}$ and $c_{Au}$ are the gold density and specific heat, $F$ is the experimental laser fluence and 
$\sigma_{abs}$ is the absorption cross section of a unit cell of the Au-NP array 
deduced from the calculations of Ref.\citenum{Magnozzi_JPCC123_16943_2019}.
This implies that interparticle interaction effects in light-absorption are taken into account.
Taking the experimental values $V=4\cdot10^3\;\mathrm{nm}^3$ and $F=(4\pm2)\; \mathrm{J/m}^2$, 
and taking $\sigma_{abs}=2\cdot10^{-16}\;\mathrm{m}^2$ \cite{Magnozzi_JPCC123_16943_2019}, we obtain $\Delta T_{max}= (75\pm30)$ K.
This value is highlighted with a blue horizontal line in Fig.~\ref{summary}(b), intercepting the thermometric calibration curve (red line) at around 20 ps time delay (blue symbol), in agreement with the experimental results of Fig.~\ref{delays}(b). Such a delay time can be interpreted as the lower limit of validity for a thermometric calibration in the considered TAS experiment, being the estimated $\Delta T_{max}$ representative of the NP temperature right after the electron-lattice equilibration.
 
To confirm this view, we resorted to numerical simulations of both thermal and TAS modulation experiments. For the plasmonic response of the nanostructures we adopted a simplified approach in which the individual nanoparticles are treated as isolated oblate nanoellipsoids in the quasi-static limit in a homogeneous environment with effective refractive index $n_{e}$~\cite{Silva_PRB_2018}. To mimick the red shift and broadening of the plasmonic resonance arising from interparticle couplings and inhomogeneities along the sample, we took $n_{e}$ and the RT Drude damping constant of gold permittivity, $\Gamma$, as fitting parameters. The optical nonlinearity of gold was modeled according to the same approach reported in previous papers, including electronic and lattice heating effects being responsible to the subsequent modulation of interband and intraband permittivity contributions, respectively (see, {\em e.g.}, Ref.~\citenum{Zavelani_ACSPhoton} and references therein for details). For the simulations of the static thermomodulation spectra, we simply assumed $T_l = T_e = T_{bath}$, whereas for the TAS simulations we employed the so-called three-temperature model~\cite{Zavelani_ACSPhoton}, to properly account for the electron and lattice temperature dynamics induced by pump absorption. The simulations confirm the general trend observed in the experiments in terms of a strong mismatch between the thermo-modulation spectra at high temperatures and the TAS spectra  for moderately low pump fluence (around 2~J/m$^2$) (Fig.~\ref{delays}(d)). Under this excitation condition, a good matching is instead retrieved between thermo-modulation spectra at intermediate temperatures and TAS spectra at around 10-20 {\em ps} time delay (Fig.~\ref{delays}(e)). \\The simulations enabled us to explore the validity of a thermometric calibration also for very high pump fluences, approaching the damage threshold. Note that for 12~J/m$^2$ fluence the TAS spectrum well overlaps with the thermo-modulation spectrum at the highest temperature (660 K), provided that a time delay of 40-50 {\em ps} is considered (Fig.~\ref{delays}(f)). Such a scaling of the time delay lower limit in the thermometric calibration is evidently connected to the electron-lattice equilibration which is expected to take more time for increasing pump fluence and thus for increasing $\Delta T_{max}$~\cite{Hodak_CPL_1998}. This behavior is well captured by the numerical simulations of Fig.~\ref{summary}(c), where we have tracked the onset of electron-lattice equilibration by defying the time delay $\tau_{EQ}$ upon which $\Delta T_e(\tau_{EQ}) - \Delta T_l(\tau_{EQ}) \simeq 1\times 10^{-2}\cdot \Delta T_l(\tau_{EQ})$ (diamonds in Fig.~\ref{summary}(c)).

The calibration procedure reported here is universal in the sense that its physical background lies in the photo-thermal response of Au, which is characteristic of the material only. 
Therefore, when changing the geometrical configuration of the nanostructures, both thermo-modulation and TAS spectra are modified in the same way, at least after {\em e-ph} thermalization has occurred. 
The calibration reported in Fig. \ref{summary}(b) is however not universally quantitatively exploitable because, whereas there is a unique way to perform thermo-modulation experiments, the heat deposited in a nanostructure will critically depend on its geometrical and dielectric parameters. 
Thus, for a given nanostructure, different pump wavelength or fluence, different quantities of heat can be deposited in the system, thereby modifying the electron heat capacity and the {\em e-ph} thermalization time.\\
The agreement between the calibration curve and the temperature-rise estimate based on Eq. 1 implies that the amount of heat deposited in the NPs can be correctly assessed as the product of the fluence by the absorption cross section. According to our experimental values, we can therefore estimate the energy deposited on a single NP as 
$F\cdot\sigma_{abs}=(8\pm4)\cdot10^{-16}$ J. 
Our method can be therefore fruitfully exploited to determine the amount of heat deposited in a nanostructure, a quantity often subject to large experimental uncertainties, albeit within the above-described general limits of applicability
of the thermo-modulation calibration.

\section*{Conclusion}

Summarizing, we have reported an approach for experimentally extracting a thermometric calibration scale
of ultrafast processes based on the comparison of the dynamic plasmonic response of
impulsively-excited Au plasmonic NPs with the corresponding static response as a function of temperature.
Exploiting this method, we experimentally retrieved the evolution of the NP
temperature as a function of the delay time $\tau$.
The results obtained with such a thermometric calibration were in good agreement with estimates of the NP temperature based on an independent model, confirming the soundness of the approach.\\
The method reported here is conceptually general, and applicable to all systems for which it is experimentally conceivable to measure
the optical response following both ultrafast excitation and, separately, as a function of externally-controlled temperature.
The application to plasmonic particles discussed here benefits from the resonant sensitivity of the optical response to
the physical parameters of the material \cite{Kelly_JPCB107_668_2003}, 
but the method is not limited to this kind of system.
The thermometric calibration that we deduce provides a $T_{e,l}(\tau)$ curve that represents a 
highly-relevant input for all theoretical models of ultrafast relaxation dynamics and may help
to explore fine details of the materials response to temperature under off-equilibrium conditions.

\section*{Supporting Information}

Atomic force image of the Au NP array. Description of the experimental apparatus. Numerical model for the calculation
of the static and dynamic plasmonic response. Comparison between experimental and calculated static and dynamic plasmonic response.

\section*{Acknowledgments}

We acknowledge support from the Ministero dell'Istruzione, dell'Universit\`{a} e della Ricerca under the PRIN Grant 2015CL3APH and
from the Compagnia di San Paolo (proj. PanLab). This project has received funding from the European Union's Horizon 2020 research and innovation programme under the Marie Sk\l odowska-Curie grant agreement N$^{\circ}$799126. G.D.V. acknowledges partial support from the Ministero dell'Istruzione dell'Universit\`{a} e della Ricerca under the PRIN Grant 2015WTW7J3.


\providecommand{\latin}[1]{#1}
\makeatletter
\providecommand{\doi}
  {\begingroup\let\do\@makeother\dospecials
  \catcode`\{=1 \catcode`\}=2 \doi@aux}
\providecommand{\doi@aux}[1]{\endgroup\texttt{#1}}
\makeatother
\providecommand*\mcitethebibliography{\thebibliography}
\csname @ifundefined\endcsname{endmcitethebibliography}
  {\let\endmcitethebibliography\endthebibliography}{}

\newpage

\newpage

\begin{center}
For table of Contents Use Only

\vspace{0.4cm}

{\LARGE Thermometric Calibration of the Ultrafast Relaxation Dynamics in Plasmonic Au Nanoparticles}
\vspace{1cm}

	{\normalsize Marzia Ferrera, Giuseppe Della Valle, Maria Sygletou, Michele Magnozzi, Daniele Catone, Patrick O'Keeffe, Alessandra Paladini, Francesco Toschi, Lorenzo Mattera, Maurizio Canepa, Francesco Bisio}\\

\vspace{0.8cm}
\end{center}

We compare the dynamic optical response of plasmonic Au nanoparticles following ultrashort-pulse
excitation with their static optical response as a function of increasing temperature.
From this comparison, the optical fingerprints of equilibrium or off-equilibrium state can be clearly identified, allowing to extract a dynamic thermometric calibration scale
of the relaxation process.

\vspace{0.8cm}

\begin{center}
\begin{figure}[!htbp]
\includegraphics[width=8.9cm]{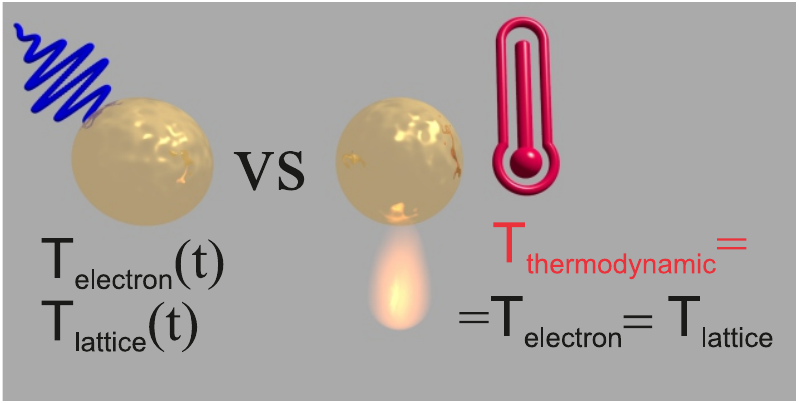}
\caption*{Graphical Table of Contents}
  \label{toc}
\end{figure}

\end{center}

\end{document}